\documentclass[preprint,authoryear,12pt]{elsarticle}
\usepackage{amssymb}

\journal{New Astronomy}

\begin{document}

\begin{frontmatter}

\title{2MASS J01074282+4845188: a new nova-like cataclysmic star with a deep eclipse}

\author[label2]{Dinko P. Dimitrov\corref{cor1}}
\ead{dinko@astro.bas.bg}
\cortext[cor1]{Corresponding author}
\author[label3]{Diana P. Kjurkchieva}
\ead{d.kyurkchieva@shu-bg.net}

\address[label2]{Institute of Astronomy and National Astronomical Observatory, Bulgarian Academy of Sciences, Tsarigradsko shossee 72, 1784 Sofia, Bulgaria}
\address[label3]{Department of Physics, Shumen University, 9700 Shumen, Bulgaria}

\begin{abstract}
We present $VRI$ photometry and low-resolution spectroscopy of the
object 2MASS J01074282+4845188. The V-shape of the eclipse, the
phase variability of the colour indices as well as the presence of
a pre-eclipse hump, standstill and flickering allow us to conclude
that it is a nova-like cataclysmic star. This is
supported by the observed broad emission H$\alpha$ line. Its
single profile with a relatively narrow FWHM but large FWZI is typical
for a nova-like variable of SW Sex subtype. The observed deep eclipses make
the newly discovered cataclysmic star 2MASS J01074282+4845188 an
interesting object for future investigation.
\end{abstract}

\begin{keyword}
binaries: eclipsing \sep binaries: close \sep cataclysmic variables \sep
stars: individual: 2MASS J01074282+4845188
\end{keyword}

\end{frontmatter}

\section{Introduction}

Many large photometric surveys were recently published:
OGLE-III \citep{udalski03}, ASAS \citep{pojmanski02},
NSVS \citep{wozniak04}, HATnet \citep{bakos04}, TrES
\citep{alonso04}, SuperWASP \citep{pollacco06}, \textit{CoRoT}
\citep{baglin02}, \textit{Kepler}
 \citep{borucki10}, etc. After
reaching the survey's original purpose, the huge photometric
data-sets collected by these surveys were released for the
community and allowed to extract (as a by-product) a great number
of light curves of known and newly discovered variable stars. Some
of them were objects of study using automated pipelines but only a
small part of them were targets of follow-up by precise observations
and analysis.

Searching for binary systems for follow-up observations with periods below 0.23 d in the
photometric surveys we noted the object T-And0-10518 in the list
of 773 eclipsing binaries \citep{devor08} found on the basis of the
TrES data. It was classified as "ambiguous EB" with coordinates
$\alpha$=01$^{\rm h}07^{\rm m}44^{\rm s}.417$ and
$\delta$=+$48^\circ 44^\prime 58^{\prime\prime}.11$. Its period of
0.1935761 d \citep[see table 6 of][]{devor08} fulfills our
criterion for a short-period binary appropriate for follow-up
observations. We began these at the beginning of 2011 and found this binary
to be a new nova-like system.

The paper presents the results from this study.

\section{Observations}

Our CCD photometric observations in $VRI$ bands were carried out
at Rozhen National Astronomical Observatory with the 60-cm
Cassegrain telescope using the FLI PL09000 CCD camera (3056 x 3056
pixels, 12 $\mu$m/pixel, field of 27.0 x 27.0 arcmin with focal
reducer). The average photometric precisions per data point in B, V
and I bands are 0.020, 0.010 and 0.008 mag respectively.

The spectra of the target were obtained by the 2-m RCC telescope
equipped with VersArray CCD camera (512 $\times$ 512 pixels, 24 $\mu$m
pixel, field of 6 $\times$ 6 arcmin) and focal reducer FoReRo-2.
The resolution of our spectra is 5.223 \AA/pixel and they cover
the range 5000--7000 \AA~using a grism with 300 lines/mm.

Table \ref{tab:log1} presents the journal of our observations.

\begin{table}
\begin{minipage}[t]{\textwidth}
\caption{Journal of the photometric and spectroscopic
observations} \label{tab:log1} \centering
\begin{footnotesize}
\renewcommand{\footnoterule}{}
\begin{tabular}{ccrrc}
\hline
Date  & Filter   & Exp. [s] &  N  & Telescope \\
\hline
2011 Jan 29 & $V, I$    & 120, 120    & 67, 67  & 60-cm \\
2011 Jan 30 & $V, I$    & 120, 120    & 65, 65  & 60-cm \\
2011 Jan 31 & $R$       & 60          & 222     & 60-cm \\
2011 Feb 07 & spectra   & 300         & 4       &  2-m  \\
\hline
\end{tabular}
\end{footnotesize}
\end{minipage}
\end{table}

\begin{table}
\begin{minipage}[t]{\textwidth}
\caption{Colours of the target and standard stars}
\label{tab:colours} \centering
\begin{scriptsize}
\renewcommand{\footnoterule}{}
\begin{tabular}{l c c c c c c}
\hline
 Star  &     ID      & $V$   & $B-V$ & $V-R$ & $V-I$ & $J-K$ \\
       &GSC1.2(2.3)  & [mag] & [mag] & [mag] & [mag] & [mag] \\
\hline
Var & NBVX029228  & 14.94 &  0.22 & -0.01 &  0.48 &  0.46 \\
St1 & 3268-0055   & 11.98 &  0.20 &  0.08 &  0.64 &  0.19 \\
St2 & 3268-0507   & 12.95 &  0.84 &  0.47 &  1.42 &  0.20 \\
St3 & 3268-0083   & 13.16 &  0.28 &  0.14 &  0.74 &  0.35 \\
St4 & 3268-0735   & 12.79 &  0.74 &  0.24 &  0.94 &  0.39 \\
St5 & 3267-1164   & 12.77 &  0.36 &  0.14 &  0.73 &  0.25 \\
St6 & 3268-0459   & 12.95 &  0.49 &  0.21 &  0.88 &  0.29 \\
St7 & 3272-0364   & 12.28 &  0.31 &  0.14 &  0.72 &  0.25 \\
St8 & NBVX029156  & 13.73 &  0.47 &  0.19 &  0.82 &  0.29 \\
\hline
\end{tabular}
\end{scriptsize}
\end{minipage}
\end{table}

\begin{figure}
 \centering
 \includegraphics[width=0.5\textwidth,bb=0 0 400 400,keepaspectratio=true]{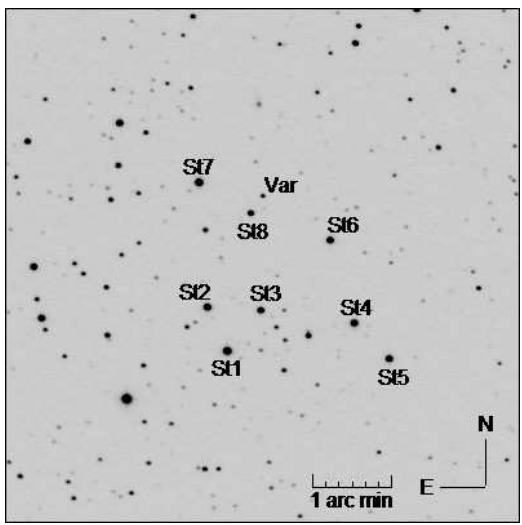}
 \caption{The field with the true variable marked as Var and T-And0-10518 marked as St8}
 \label{fig:field}
\end{figure}

The standard \textsc{idl} procedures (adapted from
\textsc{daophot}) were used for reduction of the photometric data.
For transition from instrumental system to standard photometric
system we used standard stars of \citet{landolt92} and standard
fields of \citet{stetson00}. The standard stars in the observed
field (Fig. \ref{fig:field}) were chosen by the criterion to be
constant within 0.005 mag during the all observations and in all
filters. Table 2 presents their colours $V$, $B-V$, $V-R$ and $V-I$
determined by our observations (for the target they correspond to
phase 0.25 of the out-of-eclipse light) as well as $J-K$ colours
from the 2MASS catalog \citep{skrutskie06}.

Our photometric observations revealed that the coordinates of
T-And0-10518 given by \citet{devor08} belong to the non-variable
star 2MASS J01074441+4844581 marked by St8 in Fig.
\ref{fig:field}. It turned out that the true variable star (marked
with Var in Fig. \ref{fig:field}) with a period of around 0.19 d
is the object 2MASS J01074282+4845188. It is at a distance of nearly
26 arcsec from St8 and it is around 3 times weaker than St8. To
escape the future misunderstanding we will use further the name
2MASS J01074282+4845188 for the target.

With the periodogram analysis of all our photometric data
performed by using the \textsc{persea} software (written by G.
Maciejewski, www.astri.uni.torun.pl/\\$\sim$gm/software.html)
based on ANOVA technique \citep{sc96} we derived period of
0.1935972 d. This value is slightly bigger than that in the table
6 of \citet{devor08}. Actually, their table contains two periods:
0.1935761 d (assuming an unseen secondary eclipse) and the double
value 0.3871522 d (assuming equal eclipses).

\begin{figure}
 \centering
 \includegraphics[width=0.5\textwidth,bb=0 0 340 340,keepaspectratio=true]{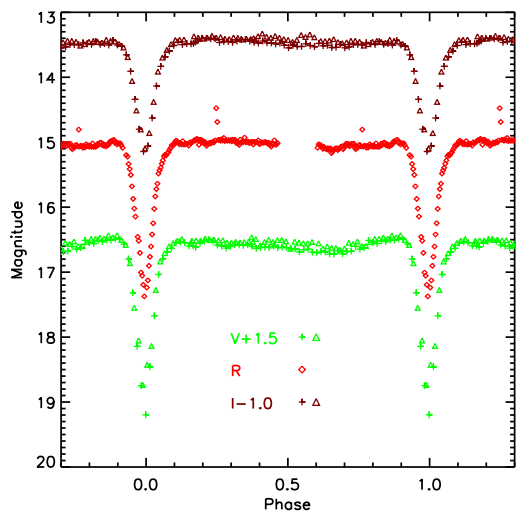}
 \caption{The $VRI$ light curves of 2MASS J01074282+4845188 from Jan 2011. The used
symbols are: pluses for the data from 2011 Jan 29; triangles for
the data from 2011 Jan 30, and diamonds for the data from 2011 Jan
31.}
 \label{fig:rozhen}
\end{figure}

The Rozhen' photometric observations were phased according to the
period derived by us. Figure \ref{fig:rozhen} presents the
corresponding folded curves in $V$, $R$, and $I$ filters.

\begin{figure}
 \centering
 \includegraphics[width=0.5\textwidth,bb=0 0 340 226,keepaspectratio=true]{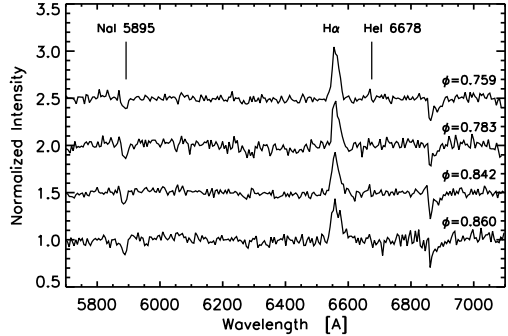}
 \caption{The low-resolution spectra of 2MASS J01074282+4845188 from 2011 Feb 07}
 \label{fig:spectra}
\end{figure}

The reduction of the spectra was performed using \textsc{iraf}
packages by bias subtraction, flat fielding, cosmic ray removal,
one-dimensional spectrum extraction and wavelength calibration.
The spectra of the target are presented in Fig. \ref{fig:spectra}.

\begin{figure}
 \centering
 \includegraphics[width=0.5\textwidth,bb=0 0 340 340,keepaspectratio=true]{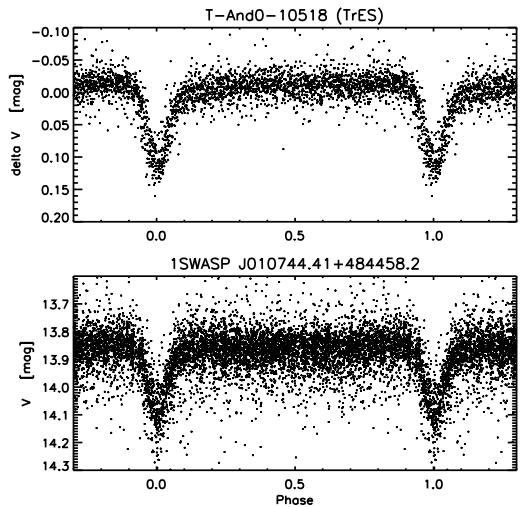}
 \caption{Top: The TrES light curve of T-And0-10518 (original data);
Bottom: Light curve from the SuperWASP photometry}
 \label{fig:tres}
\end{figure}

In order to compare our photometry with the previous one we
exhibit in Fig. \ref{fig:tres} (top panel) the TrES data (phase,
differential magnitude) of T-And0-10518 from the VizieR database.

Moreover, we checked if there are photometric observations of
2MASS J01074282+4845188 in the SuperWASP data-set
\citep{butters10}. The result was negative. However, using for
identification the coordinates of 2MASS J01074441+4844581 (or
T-And0-10518, named in the SuperWASP as 1SWASP
J010744.41+484458.1) we managed to download the photometric data of which
periodogram analysis led to the ephemeris

\begin{equation}\label{eq:1}
 HJD(MinI)=2454417.382244 + 0.1935980\times E .
\end{equation}

The light curve (Fig. \ref{fig:tres} bottom panel) on the
SuperWASP data looks the same as the TrES light curve and their
periods are almost equal. This means that the SuperWASP survey has
made the same misidentification of the variable star as the TrES
survey probably due to the same reasons: low spatial resolution of the
observations causing considerable blending from the nearby
brighter star and automatic reduction of data.

\section{Analysis of the Rozhen' observations}

\subsection{Analysis of the new photometric data}

The Rozhen' light curve of 2MASS J01074282+4845188 has one
deep, asymmetric, V-shaped, light minimum. Such type of minima are
typical for the cataclysmic nova-like stars and they are
attributed to eclipses of their accretion disks.

We found several peculiarities of the Rozhen' light curve of the target
that support our suspicion that this system is nova-like.

(a) There is a standstill on the increasing branch of the light
minimum. Such a feature is present on the light curves of eclipsing
cataclysmic stars and it is attributed to hot spot on their
accretion disks.

(b) There is a pre-eclipse hump (most visible in V colour) on the
Rozhen light curve, another feature typical for the eclipsing
cataclysmic stars, that is also attributed to the emission of the
hot spot.

(c) There is broad and shallow light decrease on the Rozhen'
curves (especially in $V$ colour) centered at phase 0.65 (Fig.
\ref{fig:rozhen}) shape and depth of which are variable. Such a
feature is hardly visible on the SuperWASP light curve.

(d) The colour index $V-R$ changes during the cycle reaching
an extreme value of $-0.01$ mag for the out-of-eclipse light (phase
0.25) and $0.33$ mag for the minimum light (phase 0.0). The colour
index $V-I$ reaches an extreme value of $0.48$ mag for the
out-of-eclipse light (phase 0.25) and $1.56$ mag for the minimum
light (phase 0.0). Such a variability of the colour indices
indicates that the star becomes bluer out of the eclipse and
redder in the middle of the eclipse. This means that the emission
of the eclipsed component at phase 0.0 (probably an accretion disk
covered by a late-type star) is strongest in $V$ band.

\begin{figure}
 \centering
 \includegraphics[width=0.5\textwidth,bb=0 0 340 226,keepaspectratio=true]{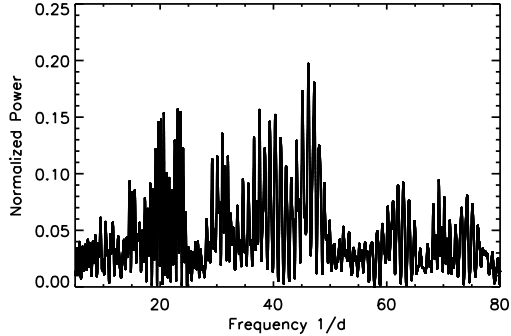}
 \caption{Periodogram analysis of 2MASS J01074282+4845188}
 \label{fig:fourie}
\end{figure}

(e) In order to search for flickering, another important
characteristic of the CVs, we made periodogram analysis of the
Rozhen' photometric data. For this aim we excluded the points of
the deep light minima (between phases 0.91 and 1.09). Moreover,
the out-of-eclipse data were de-trended in order to escape a
variability related with the orbital period (light decrease around
the phase 0.65). As a result we obtained the periodogram shown in
Fig. \ref{fig:fourie}. The figure shows variability at many frequencies,
as is typical for flickering.

Hence, our target has got all important photometric appearances of
cataclysmic nova-like eclipsing variable. Its period' value is
also appropriate for CVs.

Using the empirical relation "period -- absolute magnitude" for
the nova-like stars \citep{warner95} we obtained for 2MASS
J01074282+4845188 absolute magnitude $M_{V}$=4.5 mag and distance
$D= 1260$ pc.

\subsection{Analysis of the spectra}

The Rozhen' low-resolution spectra of 2MASS J01074282+4845188
(Fig. \ref{fig:spectra}) reveal that the most noticeable spectral
feature in the observed range is the H$\alpha$ line which profile
is typical for a cataclysmic star.

(a) The H$\alpha$ line is in emission. Its big width corresponds
to the high rotational velocity of an accretion disk.

(b) The profile of the H$\alpha$ line changes rapidly in the
framework of our short observational run (see Fig.
\ref{fig:spectra} and Table 3): (i) the EW varies between 11.4 and
13.8 \AA; (ii) the FWHM ranges between 23.7 and 33.5 \AA; (iii)
the FWZI varies between 58 and 84 \AA; (iv) the normalized
intensity changes between 1.43 and 1.55. Such parameters of the
H$\alpha$ line are typical for the accretion disks of CVs. The
observed fast changes of the H$\alpha$ profile may due to
inhomogeneity of the accretion disk.

(c) The H$\alpha$ profile is asymmetric which may be explained by
asymmetry of the accretion disk.

(d) H$\alpha$ shows an "inverse P Cyg" profile at phases 0.783 and
0.842 as well as double-peaked core at phase 0.86 at which its
FWHM increases considerably.

(e) The H$\alpha$ line is blue-shifted by around 125 km/s (see the
third column of the Table 3 containing the wavelength
$\lambda_{c}$ of the center of the profile). Taking into account
that our spectral observations were around the second quadrature
of the binary we attributed this Doppler shift to the orbital
motion of the H$\alpha$ emission source.

Besides the emission in the H$\alpha$ line, weak emission is visible
in the HeI 6678 line on some of our spectra (Fig.
\ref{fig:spectra}).

The foregoing results should be considered as preliminary due
to the low S/N of the presented spectral observations and their
poor phase covering.

\begin{table}
\begin{minipage}[t]{\textwidth}
\caption{Parameters of the H$\alpha$ line} \label{tab:colours}
\centering
\begin{scriptsize}
\renewcommand{\footnoterule}{}
\begin{tabular}{l c c c c c c c}
\hline
spectrum&   HJD  & phase & $\lambda_{c}$& FWHM & FWZI & EW & Intensity \\
       &2455000+ &       & [\AA]  & [\AA] & [\AA]& [\AA] & \\
\hline

1 &  600.21450  & 0.759 &  6559.7 &  24.9 & 68.0 &  13.8 & 1.55  \\
2 &  600.21919  & 0.783 &  6560.0 &  23.7 & 57.5 &  11.8 & 1.47  \\
3 &  600.23052  & 0.842 &  6559.1 &  26.8 & 73.2 &  11.4 & 1.43  \\
4 &  600.23403  & 0.860 &  6561.4 &  33.5 & 83.7 &  13.8 & 1.43  \\

\hline
\end{tabular}
\end{scriptsize}
\end{minipage}
\end{table}

\subsection{The subtype classification of 2MASS J01074282+4845188}

The criteria for the subtype classification of the nova-like stars
are not firm and a given star may belong to several
subtypes \citep{warner95}. We tried to determine the subtype of
2MASS J01074282+4845188 on the basis of several considerations.

(a) The presence of broad H$\alpha$ emission excludes its
classification as a UX UMa subtype.

(b) The single-peaked shape of the H$\alpha$ emission line and its
fast variability are typical for SW Sex-subtype stars.

(c) The FWHM and FWZI of the H$\alpha$ line fall into the ranges
for SW Sex stars for which Balmer and HeI emission lines have a relatively
narrow FWHM (around 1000 km s$^{-1}$) but large FWZI (2000-3000 km
s$^{-1}$).

(d) The SW Sex stars are intrinsically very luminous due to high
mass transfer rate. 2MASS J01074282+4845188 is also a quite luminous
source ($M_{V}$=4.5 mag).

(e) Although the orbital period of 4.65 h of the target is
slightly above the period' range 3 -- 4.5 h for most SW Sex
subtype stars it does not exclude the SW Sex-subtype
classification of 2MASS J01074282+4845188 because there are
several other exceptions from this criterion (see table 6 in
\citet{rodr07}).

Hence, the Rozhen' data directed us to SW Sex-subtype
classification of 2MASS J01074282+4845188. Due to the lack of
prolonged photometric observations and high-resolution spectra we
are not able to check the further criteria of this subtype
classification.

The depth of the light minimum of the Rozhen curves (Fig.
\ref{fig:rozhen}) is 2.7 mag in $V$ band while the light
amplitudes of TrES and SuperWASP light curves (Fig.
\ref{fig:tres}) are considerably smaller. The situation is very
similar to the false positives of exoplanet' candidates in the
TrES survey \citep[for instance exoplanet candidate T-Cyg1-14777, see
Fig. 5 in][]{dimitrov09}.

It should be noted that among around 2000 known cataclysmic stars
only several hundred are nova-like and only several tens of them
are eclipsing. Usually the eclipse depths of nova-like variables
are $\leq$ 1 mag. The deepest eclipses belong to the SW Sex' stars. Only nine SW Sex'
stars in the list of \citet{rodr07} have eclipse depths above 2.0
mag but smaller than that of 2MASS J01074282+4845188 (2.7
mag). Deeper eclipses of 3.2 -- 3.4 mag have been registered for
two SW Sex' stars, DW UMa \citep{stanishev04} and V1315 Aql
\citep{papadaki09}, but only during time intervals of 2 -- 3 days
whereas normally their eclipse depths are below 2.0 mag. Thus our
study not only added a new member to the small family of eclipsing
nova-like stars but 2MASS J01074282+4845188 turned out to have one of
the deepest eclipses. It is possible for the big depth of its eclipse
to be transient effect as in the cases of DW UMa and V1315 Aql.

\section{Conclusion}

The main results of our investigation are:

\begin{description}
\item[(1)] The V-shape of the eclipse, the phase behaviour of the
colour indices as well as the presence of standstill, pre-eclipse
hump and flickering allow us to conclude that 2MASS
J01074282+4845188 is a nova-like cataclysmic variable.

\item[(2)] The broad emission H$\alpha$ line of 2MASS
J01074282+4845188 is typical for a fast-rotating accretion disk of
the nova-like cataclysmic stars. The single H$\alpha$ profile with a
relatively narrow FWHM but large FWZI is typical for the SW Sex
subtype of the nova-like variables.

\item[(3)] The deep eclipses observed at the beginning of 2011
make the newly discovered cataclysmic star 2MASS J01074282+4845188
an interesting object for future observations and investigation.

\end{description}

\section*{Acknowledgments}

The research was supported partly by funds of project DO 02-362 of
the Bulgarian Ministry of Education and Science. This research
make use of the SIMBAD and VizieR databases, operated at CDS,
Strasbourg, France, and NASA's Astrophysics Data System Abstract
Service.

This research was based on data collected with
the telescopes at Rozhen National Astronomical Observatory.
Authors gratefully acknowledge observing grant support from the
Institute of Astronomy and Rozhen National Astronomical
Observatory, Bulgarian Academy of Sciences.

The authors are thankful to the anonymous referee for the useful notes and advices.

\end{document}